# Nature of isoscalar dipole resonances in heavy nuclei

V.I. Abrosimov[1], O.I. Davidovskaya

Institute for Nuclear Research, Nat. Acad. Sci. of Ukraine

*Isoscalar dipole nuclear response reveals the low- and high-energy resonances. The nature of isoscalar dipole resonances in heavy spherical nuclei is studied by using the translation-invariant kinetic model of small oscillations of finite Fermi systems. Calculations of the velocity field at the centroid energy show a pure vortex character of the low-energy isoscalar dipole resonance in spherical nuclei and confirm the anisotropic compression character of the high-energy one. The evolution of the velocity field in dependence on the excitation energy of the nucleus within the resonance width is studied. It is found that the low-energy isoscalar dipole resonance retains a vortex character while with increasing energy this collective excitation also involves compression. The high-energy resonance keeps the compression character with a change in the excitation energy within the resonance width but the compression-expansion region of the velocity field related to this resonance shifts inside the nucleus.*

**1. Introduction**

Recent experiments have shown that the isoscalar dipole nuclear response reveals not only the compression dipole mode (the high-energy resonance) but also a low-energy resonance [1-4]. Theoretical studies of the nuclear isoscalar dipole excitations have been carried out within quantum approaches [5-11] as well as using semi-classical ones [12-15]. It was found that the low-energy isoscalar dipole resonance has an essentially vortex character [16-19]. It is of interest to further study the nature of this resonance. For this purpose it is suitable to consider the velocity field related to the resonance. This local dynamic quantity describes the spatial distribution of the average nucleon velocity in collective excitation and provides information on the nature of the excitation.

In this paper the collective isoscalar dipole excitations are considered within the translation-invariant kinetic model of small oscillations of finite Fermi systems [20]. This model makes it possible to find an analytic expression for the dynamic fluctuations of the distribution function in the phase space under collective isoscalar dipole excitations. By using this solution, one can obtain an explicit expression for the velocity field, which is a local quantity that depends on the excitation energy of the system. In Ref. [21], the isoscalar dipole response function was studied within the kinetic model [20], taking into account a residual interaction between nucleons. It was found that the response function has two-resonance structure and reproduces the experimental values for the centroid energies of the low- and high-energy isoscalar dipole resonances in heavy nuclei. We will use this response function in the present paper.

---

[1] Corresponding author.
*E-mail address:* abrosim@kinr.kiev.ua

In Section 2, the analytic expression of the velocity field related to the isoscalar dipole excitations is presented within the translation-invariant kinetic model of small oscillations of finite Fermi systems. To study the nature of the isoscalar dipole resonances in nuclei, the evolution of the velocity field character depending on the excitation energy of the nucleus within the resonance width is investigated (Section 3). In Appendix, the explicit expression for the dynamic fluctuations of the phase-space distribution function under the collective isoscalar dipole excitations is derived.

## 2. Velocity field in kinetic model

We consider the isoscalar dipole excitations of nuclei within a kinetic model, which is based on the Vlasov equation for finite Fermi systems with moving surface [14, 20-21]. In this model, a nucleus is treated as a gas of interacting fermions confined to a spherical cavity with moving surface. We calculate the response of nuclei to a weak external field of the kind

$$V(\vec{r},t) = \beta\ \delta(t)\ Q^{(3)}(r)\ Y_{10}(\theta,\varphi), \tag{1}$$

where $Q^{(3)}(r) = r^3$ is the second-order dipole moment, $\delta(t)$ is the Dirac delta-function in time, and $\beta$ is a parameter that describes the external field strength. Our model is the translation-invariant [14, 21] therefore the external field (1) does not excite the center of mass. We take into account the residual interaction between nucleons by using a separable effective interaction of the dipole-dipole type [21]. Within our kinetic model, we can find the explicit expression for the fluctuation of the phase-space distribution function related to the collective isoscalar dipole excitations. By using this function, we can calculate the response function [21] as well as the local dynamical quantities, in particular, the velocity field.

In our kinetic model, the time Fourier-transform of the velocity field in the linear approximation is determined as

$$\vec{u}(\vec{r},\omega) = \frac{1}{m\rho_0} \int d\vec{p}\, \vec{p}\, \delta n(\vec{r},\vec{p},\omega), \tag{2}$$

where $\vec{r},\vec{p}$ are the radius-vector and momentum of the particle, respectively; $\rho_0$ is the nuclear matter density at equilibrium and $\delta n(\vec{r},\vec{p},\omega)$ is the variation of the particle distribution function in the phase space, caused by the action of an external field (1).

Taking the Z-axis in the direction of the external field, we are interested in the velocity field in the meridian plane XZ ($\vec{r} = (x, y = 0, z)$ or in the spherical coordinates $\vec{r} = (r,\theta,\varphi = 0)$) that usually exploited in the RPA calculations [11, 22].

Then the velocity field can be written as

$$\vec{u}(r,\theta,\varphi=0,\omega)=u_z(r,\theta,\omega)\vec{e}_z+u_x(r,\theta,\omega)\vec{e}_x, \qquad (3)$$

where $u_x(r,\theta,\omega)$ i $u_z(r,\theta,\omega)$ are the projections of the velocity field vector into the X and Z axes, respectively, $\vec{e}_x, \vec{e}_z$ are unit vectors directed along these axes. For the isoscalar dipole excitations, the functions $u_x(r,\theta,\omega)$ and $u_z(r,\theta,\omega)$ have the same form as for the isovector ones, see Ref. [23]. The difference is in the external field that for the isoscalar dipole excitations is given by Eq. (1). So we get

$$u_z(r,\theta,\omega)=Y_{00}(\theta,0)u_{10}(r,\omega)-\sqrt{\frac{2}{5}}Y_{20}(\theta,0)u_{12}(r,\omega), \qquad (4)$$

$$u_x(r,\theta,\omega)=\sqrt{\frac{3}{5}}Y_{21}(\theta,0)u_{12}(r,\omega). \qquad (5)$$

Here $Y_{lm}(\theta,\varphi)$ are spherical functions. The functions $u_{10}(r,\omega)$ and $u_{12}(r,\omega)$ describe the radial dependence of the velocity field and are given by

$$u_{12}(r,\omega)=-i\sqrt{\frac{2}{3}}\pi\frac{1}{\rho_0}\frac{1}{r^2}\int d\varepsilon\int dll\sum_{N=-1}^{1}\left\{-i[\delta\tilde{n}_N^+(r,\varepsilon,l,\omega)-\delta\tilde{n}_N^-(r,\varepsilon,l,\omega)]+\right.$$
$$\left.+\frac{N}{2}\frac{l}{p(r,\varepsilon,l)r}[\delta\tilde{n}_N^+(r,\varepsilon,l,\omega)+\delta\tilde{n}_N^-(r,\varepsilon,l,\omega)]\right\}, \qquad (6)$$

$$u_{10}(r,\omega)=-i\sqrt{\frac{1}{3}}\pi\frac{1}{\rho_0}\frac{1}{r^2}\int d\varepsilon\int dll\sum_{N=-1}^{1}\left\{i[\delta\tilde{n}_N^+(r,\varepsilon,l,\omega)-\delta\tilde{n}_N^-(r,\varepsilon,l,\omega)]+\right.$$
$$\left.+N\frac{l}{p(r,\varepsilon,l)r}[\delta\tilde{n}_N^+(r,\varepsilon,l,\omega)+\delta\tilde{n}_N^-(r,\varepsilon,l,\omega)]\right\}, \qquad (7)$$

where $\varepsilon$ is particle energy, $l$ is the magnitude of its angular momentum, $p(r,\varepsilon,l)=p_F\sqrt{1-(l/p_F r)^2}$ is radial momentum; the functions $\delta\tilde{n}_N^\pm(r,\varepsilon,l,\omega)$ are the solutions of the linearized Vlasov kinetic equation for a finite system with moving surface, see Appendix.

### 3. Character of isoscalar dipole velocity fields

The numerical calculations of the dipole velocity field (4) were carried out for a Fermi system of A = 208 nucleons. In our calculations we used the following standard values of nuclear parameters: $r_0$ = 1.25 fm, $\varepsilon_F$ = 30.94 MeV, $m$ = 1.04 MeV$(10^{-22}s)^2$/fm$^2$. The parameter of the isoscalar dipole interaction strength is equal to $\kappa_1=-7.5\cdot 10^{-3}$ MeV/fm$^2$ and is determined by comparison with the giant monopole resonance data in nucleus $^{208}Pb$ within our kinetic model [14].

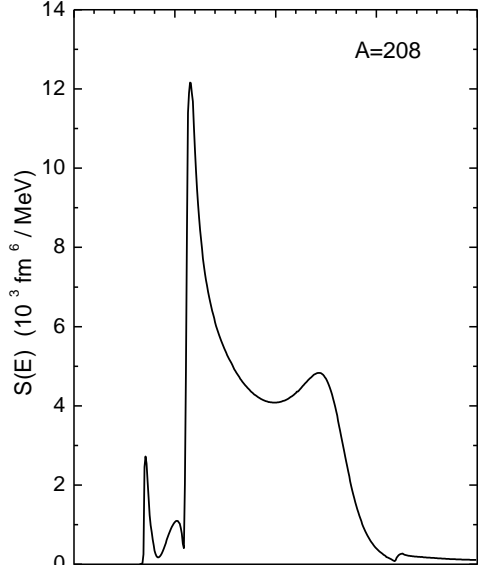

*Fig.1. Isoscalar dipole strength function for a system of A=208 interacting nucleons [21].*

In Fig.1 the isoscalar dipole strength function $S(E)$ $(E=\hbar\omega)$ is shown for a system of A=208 nucleons. This function is determined by the imaginary part of the isoscalar dipole response function calculated in Ref. [21], see Eq. (24). We can see that the isoscalar dipole strength function has a two-resonance structure. The centroid energies of the low- and the high-energy isoscalar dipole resonances equal 11.5 and 24.3MeV, respectively, in the agreement with experimental data for the isoscalar dipole resonances in $Pb^{208}$ nucleus [3, 4].

The results of numerical calculations of the velocity field (3) at the centroid energies 11.5 and 24.3 MeV corresponding to the low- and high-energy isoscalar dipole resonances (see Fig.1) are shown in Fig.2. The velocity field related to the low-energy resonance has a pure vortex character, see Fig.2a, while the velocity field for the high-energy resonance is showing the compression character, Fig.2b. The character of the velocity fields for the isoscalar dipole resonances obtained in the present paper within the kinetic (semiclassical) model agrees with the results of corresponding quantum calculations, see, e.g., [11].

In Figs. 3, 4, the evolution of the velocity field character depending on the excitation energy of the nucleus within the resonance width is shown. The velocity fields for the low-energy isoscalar dipole resonance are calculated at the excitation energies 11.2 and 12.5 MeV, see Figs. 3a and 3c. We can see that the velocity fields for the low-energy resonance keep the vortex character at the excitation energies within the resonance width. However, with increasing excitation energy the velocity field for this resonance also has the compression, see Fig. 3c. The velocity field evolution for the high-energy isoscalar dipole resonance is shown at the excitation energies 23.3 and 25.1 MeV, see Figs. 4a and 4c. This resonance preserves the compression character at the excitation energies within the resonance width but the compression-expansion region of the velocity fields shifts inside the system when changing the excitation energy.

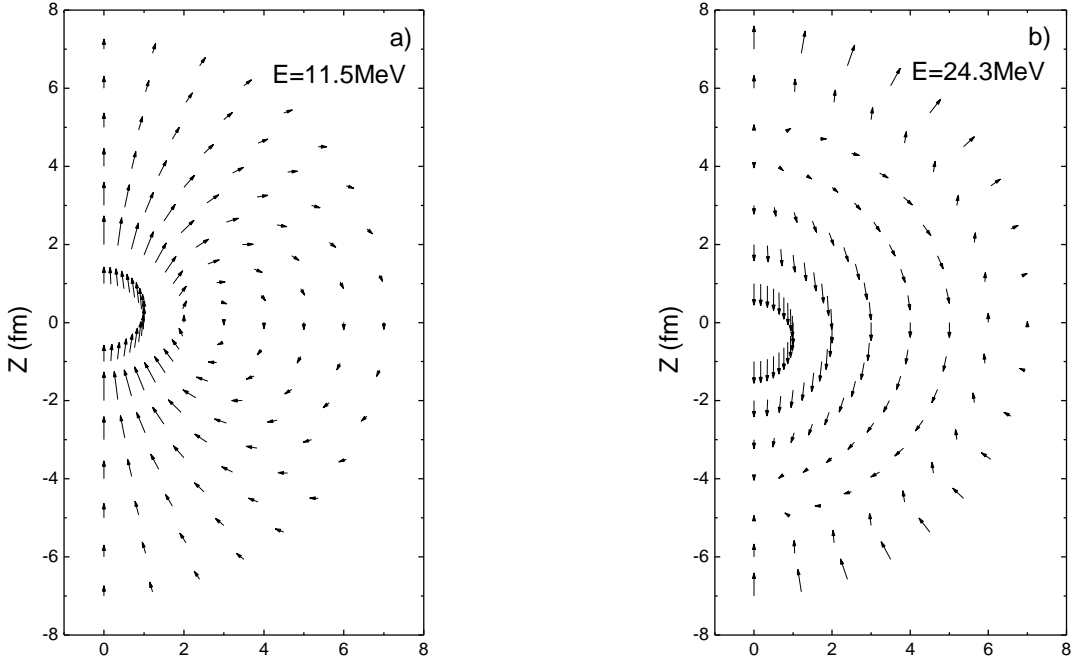

*Fig.2. Velocity fields in the XZ-plane related to the low- and high- energy isoscalar dipole resonance in a system of A=208 nucleons. The velocity fields are calculated at energies 11.5 MeV (a) and 24.3 MeV (b) which correspond to the energies of the low- and high-energy resonance peaks (the centroid energies), respectively, see Fig.1.*

### 4. Conclusions

The velocity fields related to the low- and high-energy isoscalar dipole resonances in heavy nuclei have been considered within the translation-invariant kinetic model of small oscillations of finite Fermi systems. It is found that the velocity field of the low-energy isoscalar dipole resonance at the centroid energy has an essentially vortex character. However, at increasing excitation energy of nucleus within the resonance width this collective excitation can also involve compression. Our study confirms the anisotropic ("squeezing") compression character of the high-energy isoscalar dipole resonance and discovers that the compression-expansion region of the velocity field shifts inside the nucleus with a change in the excitation energy within the resonance width.

As a final comment, we would like to add that the absence of the low-energy isoscalar dipole mode in the liquid drop model [24] gives a reason to assume that the dynamic deformation of the Fermi surface plays an important role in the formation of the low-energy resonance. To clarify this point it is of interest to study the character of the momentum flux related to this resonance.

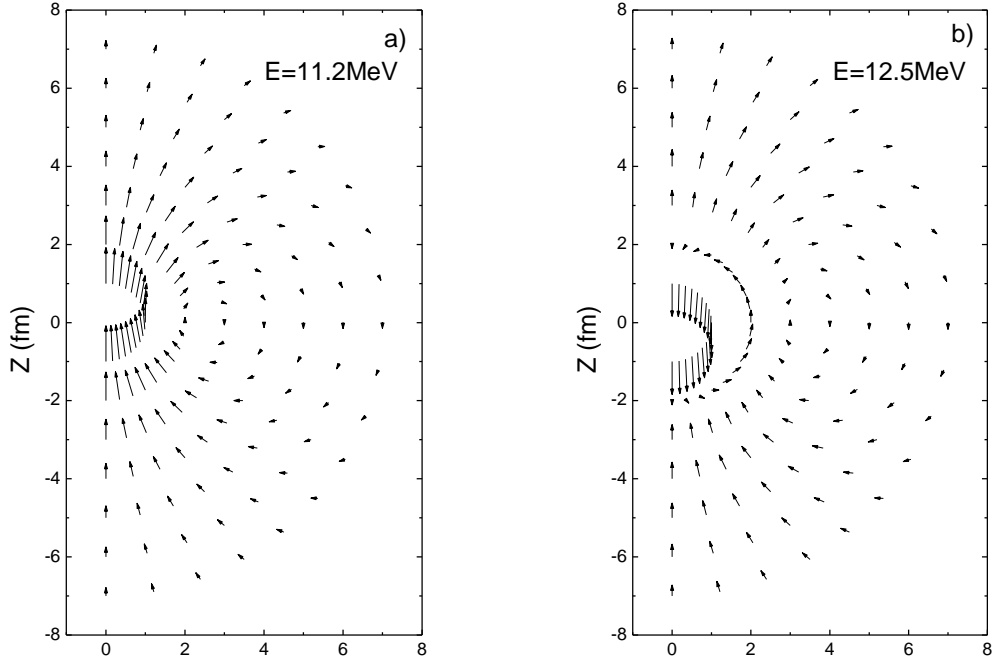

*Fig.3. Velocity fields for the low-energy isoscalar dipole resonance calculated at energies 11.2 MeV (a) and 12.5 MeV (b) within the resonance width, see Fig.1. Velocity field at the centroid energy 11.5 MeV is shown in Fig.2a. The system contains A = 208 nucleons.*

**Appendix: Fluctuations of the phase-space density for the isoscalar dipole excitations**

In this appendix, we give an explicit expression for the fluctuations of the phase-space distribution function related to collective isoscalar dipole excitations.

Solving the linearized Vlasov equation with boundary conditions at the moving surface, we can obtain an explicit expression for the fluctuations of the phase-space particle distribution functions. It can be written as [20]:

$$\delta \tilde{n}_N^\pm(r,\varepsilon,l,\omega) = \delta n_{N,3}^{0\pm}(r,\varepsilon,l,\omega)[1+\kappa_1 \tilde{R}_{13}^V(\omega)] + \delta \tilde{n}_N^{s\pm}(r,\varepsilon,l,\omega). \quad (A.1)$$

The functions $\delta n_{N,k}^{0\pm}(r,\varepsilon,l,\omega)$ in Eq. (A1) are the solutions of the Vlasov equation for a system of noninteracting nucleons confined by a fixed surface:

$$\delta n_{N,k}^{0\pm}(r,\varepsilon,l,\omega) = -\beta \frac{\partial n_0(\varepsilon)}{\partial \varepsilon} \sum_{n=-\infty}^{\infty} \omega_{nN}(\varepsilon,l) e^{\pm i[\omega_{nN}(\varepsilon,l)\tau(r,\varepsilon,l)+N\gamma(r,\varepsilon,l)]} \times$$
$$\times \frac{Q_{nN}^k(l)}{\omega - \omega_{nN}(\varepsilon,l) + i\eta}, \quad (A.2)$$

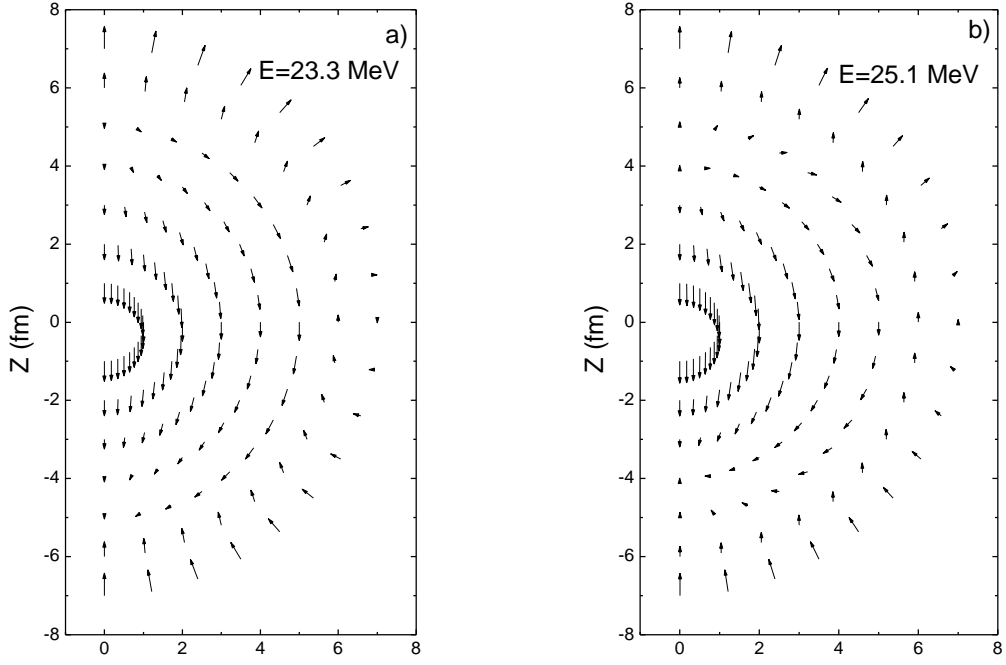

*Fig.4. Velocity fields for the high-energy isoscalar dipole resonance calculated at energies 23.3 MeV (a) and 25.1 MeV (b) within the resonance width, see Fig.1. Velocity field at the centroid energy 24.3 MeV is shown in Fig.2b. The system contains A = 208 nucleons.*

where $n_0(\varepsilon)$ is the equilibrium particle distribution function in the Thomas-Fermi approximation; $\omega_{nN}(\varepsilon,l) = n\dfrac{2\pi}{T(\varepsilon,l)} + N\dfrac{\Gamma(\varepsilon,l)}{T(\varepsilon,l)}$ are single-particle frequencies; $T(\varepsilon,l)$, $\Gamma(\varepsilon,l)$ are the periods of the radial and "angular" particle motions. In Eq. (A.2), the classical limit for the radial matrix elements of the quantum-mechanical dipole operators $r^k$ ($k=1,3$) $Q_{nN}^k(x)$ are given by

$$Q_{nN}^1(x) = (-1)^n R \frac{1}{s_{nN}^2(x)}, \qquad (A.3)$$

$$Q_{nN}^3(x) = 3R^2 Q_{nN}^1(x)\left(1 + \frac{4}{3}N\frac{\sqrt{1-x^2}}{s_{nN}(x)} - \frac{2}{s_{nN}^2(x)}\right). \qquad (A.4)$$

In Eq. (A3), (A4), we use the dimensionless variables

$$s = \frac{\omega}{v_F/R}, \quad s_{nN} = \frac{\omega_{nN}}{v_F/R}, \quad x = \sqrt{1-(l/p_F R)^2}, \quad s_{nN}(x) = \frac{n\pi + N\arcsin(x)}{x}.$$

The functions $\delta \tilde{n}_N^{s\pm}(r,\varepsilon,l,\omega)$ in Eq. (A.1) associated with the moving surface are given by

$$\delta \tilde{n}_N^{s\pm}(r,\varepsilon,l,\omega) = \frac{\partial n^0(\varepsilon)}{\partial \varepsilon} \frac{e^{\pm i[\omega \tau(r,\varepsilon,l)+N\gamma(r,\varepsilon,l)]}}{\sin[\omega T(\varepsilon,l)+N\Gamma(\varepsilon,l)]} p(R,\varepsilon,l)\omega \delta R_3(\omega). \quad (A5)$$

Here $2\tau(r,\varepsilon,l)|_{r=R} = T(\varepsilon,l)$; $2\gamma(r,\varepsilon,l)|_{r=R} = \Gamma(\varepsilon,l)$ and $\delta R_3(\omega)$ is the variation of the equilibrium radius $R$ of the system induced by the external field (1)

$$\delta R_3(\omega) = \beta \frac{R^4(\chi_3^0 + \kappa_1 \rho_0 R^3 R_{11}^0(\omega))}{-\chi_1(1-\kappa_1 R_{11}^0(\omega)) + \kappa_1 R^6(\chi_1^0 + \rho_0 R)^2}, \quad (A6)$$

where the parameter $\kappa_1$ describes the isoscalar dipole interaction strength. The functions $\chi_k^0(\omega)$ and $\chi_1(\omega)$ in Eq. (A6) describe dynamic surface effects and are given by

$$\chi_k^0(s) = \frac{9A}{8\pi} \sum_{n=-\infty}^{+\infty} \sum_{N=\pm 1} \int_0^1 dx x^2 s_{nN}(x) \frac{(-1)^n Q_{nN}^{(k)}(x)}{s+i\varepsilon - s_{nN}(x)} \quad (k=1,3), \quad (A7)$$

$$\chi_1(s) = -\frac{9A}{4\pi}\varepsilon_F(s+i\varepsilon) \sum_{n=-\infty}^{+\infty} \sum_{N=\pm 1} \int_0^1 dx x^2 \frac{1}{s+i\varepsilon - s_{nN}(x)}, \quad (A8)$$

where $\varepsilon_F$ is the Fermi energy. The function $R_{11}^0(\omega)$ in Eq. (A6) is the single-particle response function of the system confined by a fixed surface

$$R_{jk}^0(s) = \frac{9A}{16\pi} \frac{1}{\varepsilon_F} \sum_{n=-\infty}^{+\infty} \sum_{N=\pm 1} \int_0^1 dx x^2 s_{nN}(x) \frac{Q_{nN}^j(x)Q_{nN}^k(x)}{s+i\varepsilon - s_{nN}(x)}, \quad (j,k=1,3). \quad (A9)$$

Finally, the function $\tilde{R}_{13}^V(\omega)$ in Eq. (A1) is defined as

$$\tilde{R}_{13}^V(\omega) = \frac{R_s(\omega) + R_{13}^0(\omega)}{1-\kappa_1 R_{11}^0(\omega)}, \quad (A10)$$

where

$$R_s(\omega) = -\frac{1}{\beta} R^2 \delta R_3(\omega) \left[\chi_1^0(\omega) + \rho_0 R\right]. \quad (A11)$$